\overfullrule=0pt

\def\refto#1{$^{#1}$}           
\def\ref#1{ref.~#1}                     
\def\Ref#1{#1}                          
\gdef\refis#1{\item{#1.\ }}                    
\def\beginparmode{\endmode
  \begingroup \def\endmode{\par\endgroup}}
\let\endmode=\par
\def\body{\beginparmode}
\def\head#1{                    
  \goodbreak\vskip 0.5truein    
  {\centerline{\bf{#1}}\par}
   \nobreak\vskip 0.25truein\nobreak}
\def\references                 
Phys Rev
  {\head{References}            
co
mmas)\.
   \beginparmode
   \frenchspacing \parindent=0pt \leftskip=1truecm
   \parskip=8pt plus 3pt \everypar{\hangindent=\parindent}}
\def\endreferences{\body}

\catcode`@=11
\newcount\r@fcount \r@fcount=0
\newcount\r@fcurr
\immediate\newwrite\reffile
\newif\ifr@ffile\r@ffilefalse
\def\w@rnwrite#1{\ifr@ffile\immediate\write\reffile{#1}\fi\message{#1}}

\def\writer@f#1>>{}
\def\referencefile{
  \r@ffiletrue\immediate\openout\reffile=\jobname.ref%
  \def\writer@f##1>>{\ifr@ffile\immediate\write\reffile%
    {\noexpand\refis{##1} = \csname r@fnum##1\endcsname = %
     \expandafter\expandafter\expandafter\strip@t\expandafter%
     \meaning\csname r@ftext\csname r@fnum##1\endcsname\endcsname}\fi}%
  \def\strip@t##1>>{}}

\def\citeall#1{\xdef#1##1{#1{\noexpand\cite{##1}}}}
\def\cite#1{\each@rg\citer@nge{#1}}	

\def\each@rg#1#2{{\let\thecsname=#1\expandafter\first@rg#2,\end,}}
\def\first@rg#1,{\thecsname{#1}\apply@rg}	
\def\apply@rg#1,{\ifx\end#1\let\next=\relax
\else,\thecsname{#1}\let\next=\apply@rg\fi\next}

\def\citer@nge#1{\citedor@nge#1-\end-}	
\def\citer@ngeat#1\end-{#1}
\def\citedor@nge#1-#2-{\ifx\end#2\r@featspace#1 
  \else\citel@@p{#1}{#2}\citer@ngeat\fi}	
\def\citel@@p#1#2{\ifnum#1>#2{\errmessage{Reference range #1-#2\space is bad.}%
    \errhelp{If you cite a series of references by the notation M-N, then M and
    N must be integers, and N must be greater than or equal to M.}}\else%
 {\count0=#1\count1=#2\advance\count1 by1\relax\expandafter\r@fcite\the\count0,%
  \loop\advance\count0 by1\relax
    \ifnum\count0<\count1,\expandafter\r@fcite\the\count0,%
  \repeat}\fi}

\def\r@featspace#1#2 {\r@fcite#1#2,}	
\def\r@fcite#1,{\ifuncit@d{#1}
    \newr@f{#1}%
    \expandafter\gdef\csname r@ftext\number\r@fcount\endcsname%
                     {\message{Reference #1 to be supplied.}%
                      \writer@f#1>>#1 to be supplied.\par}%
 \fi%
 \csname r@fnum#1\endcsname}
\def\ifuncit@d#1{\expandafter\ifx\csname r@fnum#1\endcsname\relax}%
\def\newr@f#1{\global\advance\r@fcount by1%
    \expandafter\xdef\csname r@fnum#1\endcsname{\number\r@fcount}}

\let\r@fis=\refis			
\def\refis#1#2#3\par{\ifuncit@d{#1}
   \newr@f{#1}%
   \w@rnwrite{Reference #1=\number\r@fcount\space is not cited up to now.}\fi%
  \expandafter\gdef\csname r@ftext\csname r@fnum#1\endcsname\endcsname%
  {\writer@f#1>>#2#3\par}}

\def\ignoreuncited{
   \def\refis##1##2##3\par{\ifuncit@d{##1}%
     \else\expandafter\gdef\csname r@ftext\csname r@fnum##1\endcsname\endcsname%
     {\writer@f##1>>##2##3\par}\fi}}

\def\r@ferr{\endreferences\errmessage{I was expecting to see
\noexpand\endreferences before now;  I have inserted it here.}}
\let\r@ferences=\references
\def\references{\r@ferences\def\endmode{\r@ferr\par\endgroup}}

\let\endr@ferences=\endreferences
\def\endreferences{\r@fcurr=0
  {\loop\ifnum\r@fcurr<\r@fcount
    \advance\r@fcurr by 1\relax\expandafter\r@fis\expandafter{\number\r@fcurr}%
    \csname r@ftext\number\r@fcurr\endcsname%
  \repeat}\gdef\r@ferr{}\endr@ferences}


\let\r@fend=\endpaper\gdef\endpaper{\ifr@ffile
\immediate\write16{Cross References written on []\jobname.REF.}\fi\r@fend}

\catcode`@=12

\citeall\refto		
\citeall\ref		%
\citeall\Ref		%

\def\singlespace{\baselineskip 12pt \lineskip 1pt \parskip 2pt plus 1 pt}

\def\today{\number\day\enspace
     \ifcase\month\or January\or Febuary\or March\or April\or May\or
     June\or July\or August\or September\or October\or
     November\or December\fi \enspace\number\year}
\def\clock{\count0=\time \divide\count0 by 60
    \count1=\count0 \multiply\count1 by -60 \advance\count1 by \time
    \number\count0:\ifnum\count1<10{0\number\count1}\else\number\count1\fi}
\footline={\hss -- \folio\ -- \hss}

\def\deg{\ifmmode^\circ\else$^\circ$\fi}
\def\solar{\ifmmode_{\mathord\odot}\else$_{\mathord\odot}$\fi}
\def\jref#1 #2 #3 #4 {{\par\noindent \hangindent=3em \hangafter=1 
      \advance \rightskip by 5em #1, {\it#2}, {\bf#3}, #4.\par}}
\def\ref#1{{\par\noindent \hangindent=3em \hangafter=1 
      \advance \rightskip by 5em #1.\par}}
\newcount\eqnum
\def\nexteq{\global\advance\eqnum by1 \eqno(\number\eqnum)}
\def\lasteq#1{\if)#1[\number\eqnum]\else(\number\eqnum)\fi#1}
\def\preveq#1#2{{\advance\eqnum by-#1
    \if)#2[\number\eqnum]\else(\number\eqnum)\fi}#2}
\def\endtable{\endgroup}
\def\tableheight{\vrule width 0pt height 8.5pt depth 3.5pt}
{\catcode`|=\active \catcode`&=\active 
    \gdef\tabledelim{\catcode`|=\active \let|=\vbar
                     \catcode`&=\active \let&=\nobar} }
\def\table{\begingroup
    \def\twidth{\hsize}
    \def\tablewidth##1{\def\twidth{##1}}
    \def\defaultheight{\vrule width 0pt height 8.5pt depth 3.5pt}
    \def\heightdepth##1{\dimen0=##1
        \ifdim\dimen0>5pt 
            \divide\dimen0 by 2 \advance\dimen0 by 2.5pt
            \dimen1=\dimen0 \advance\dimen1 by -5pt
            \vrule width 0pt height \the\dimen0  depth \the\dimen1
        \else  \divide\dimen0 by 2
            \vrule width 0pt height \the\dimen0  depth \the\dimen0 \fi}
    \def\spacing##1{\def\defaultheight{\heightdepth{##1}}}
    \def\nextheight##1{\noalign{\gdef\tableheight{\heightdepth{##1}}}}
    \def\end{\cr\noalign{\gdef\tableheight{\defaultheight}}}
    \def\zerowidth##1{\omit\hidewidth ##1 \hidewidth}    
    \def\hline{\noalign{\hrule}}
    \def\skip##1{\noalign{\vskip##1}}
    \def\bskip##1{\noalign{\hbox to \twidth{\vrule height##1 depth 0pt \hfil
        \vrule height##1 depth 0pt}}}
    \def\header##1{\noalign{\hbox to \twidth{\hfil ##1 \unskip\hfil}}}
    \def\bheader##1{\noalign{\hbox to \twidth{\vrule\hfil ##1 
        \unskip\hfil\vrule}}}
    \def\spanloop{\span\omit \advance\mscount by -1}
    \def\extend##1##2{\omit
        \mscount=##1 \multiply\mscount by 2 \advance\mscount by -1
        \loop\ifnum\mscount>1 \spanloop\repeat \ \hfil ##2 \unskip\hfil}
    \def\vbar{&\vrule&}
    \def\nobar{&&}
    \def\hdash##1{ \noalign{ \relax \gdef\tableheight{\heightdepth{0pt}}
        \toks0={} \count0=1 \count1=0 \putout##1\end 
        \toks0=\expandafter{\the\toks0 &\end} \xdef\piggy{\the\toks0} }
        \piggy}
    \let\e=\expandafter
    \def\putspace{\ifnum\count0>1 \advance\count0 by -1
        \toks0=\e\e\e{\the\e\toks0\e&\e\multispan\e{\the\count0}\hfill} 
        \fi \count0=0 }
    \def\putrule{\ifnum\count1>0 \advance\count1 by 1
        \toks0=\e\e\e{\the\e\toks0\e&\e\multispan\e{\the\count1}\leaders\hrule\hfill}
        \fi \count1=0 }
    \def\putout##1{\ifx##1\end \putspace \putrule \let\next=\relax 
        \else \let\next=\putout
            \ifx##1- \advance\count1 by 2 \putspace
            \else    \advance\count0 by 2 \putrule \fi \fi \next}   }
\def\tablespec#1{
    \def\vdimens{\noexpand\tableheight}
    \def\tabby{\tabskip=0pt plus100pt minus100pt}
    \def\r{&################\tabby&\hfil################\unskip}
    \def\c{&################\tabby&\hfil################\unskip\hfil}
    \def\l{&################\tabby&################\unskip\hfil}
    \edef\templ{\noexpand\vdimens ########\unskip  #1 
         \unskip&########\tabskip=0pt&########\cr}
    \tabledelim
    \edef\body##1{ \vbox{
        \tabskip=0pt \offinterlineskip
        \halign to \twidth {\templ ##1}}} }

\newbox\grsign \setbox\grsign=\hbox{$>$}
\newdimen\grdimen \grdimen=\ht\grsign
\newbox\laxbox \newbox\gaxbox
\setbox\gaxbox=\hbox{\raise.5ex\hbox{$>$}\llap
	{\lower.5ex\hbox{$\sim$}}}\ht1=\grdimen\dp1=0pt
\setbox\laxbox=\hbox{\raise.5ex\hbox{$<$}\llap
	{\lower.5ex\hbox{$\sim$}}}\ht2=\grdimen\dp2=0pt

\def\uJy{\ifmmode{\,\mu{\rm Jy}}\else$\,{\mu{\rm Jy}}$\fi}
\def\mJy{\ifmmode{\,{\rm mJy}}\else${\,{\rm mJy}}$\fi}
\def\MHz{\ifmmode{\,{\rm MHz}}\else{$\,{\rm MHz}$}\fi}
\def\GHz{\ifmmode{\,{\rm GHz}}\else{$\,{\rm GHz}$}\fi}
\def\solar{\ifmmode_{\mathord\odot}\else$_{\mathord\odot}$\fi}
\def\Msolar{\ifmmode{\, {\rm M\solar}}\else{${\, {\rm M\solar}}$}\fi}
\def\Rsolar{\ifmmode{\, {\rm R\solar}}\else{${\, {\rm R\solar}}$}\fi}
\def\kms{\ifmmode{\,{\rm km\,s^{-1}}}\else${\,{\rm km\,s^{-1}}}$\fi}
\def\kpc{\ifmmode{\,{\rm kpc}}\else${\,{\rm kpc}}$\fi}
\def\us{\ifmmode{\,\mu{\rm s}}\else$\,{\mu{\rm s}}$\fi}
\def\ms{\ifmmode{\,{\rm ms}}\else$\,{{\rm ms}}$\fi}
\def\y{\ifmmode{\,{\rm y}}\else$\,{\rm y}$\fi}
\def\h{\ifmmode{^{\rm h}}\else$^{\rm h}$\fi}
\def\m{\ifmmode{^{\rm m}}\else$^{\rm m}$\fi}
\def\s{\ifmmode{^{\rm s}}\else$^{\rm s}$\fi}
\def\Lmin{\ifmmode{L_{min}}\else{$L_{min}$}\fi}

\input psfig.sty

\magnification=\magstep1
\singlespace

\font\eightrm=cmr8



\font\lgh=cmbx10 scaled \magstep2
\def\sgr{SGR\thinspace 1900+14}
\def\snr{G\thinspace 42.8+0.6}
\def\xray{RX~J190717+0919.3}
\def\vla{VLA~J190714.3+091920}
\def\hb{\hfill\break}

\def\localization{1}
\def\radio{2}

\def\fluxtable{1}


\smallskip
\hrule
\bigskip
\line{\lgh A relativistic particle outburst from the \hb}
\line{\lgh soft gamma-ray repeater 1900+14 \hb}

\bigskip

\bigskip

\line{D.~A.~Frail$^1$, S.~R.~Kulkarni$^2$ \&\ J.~S.~Bloom$^2$ \hb}

\bigskip

\line{$^1$National Radio Astronomy Observatory, P.~O.~Box 0,
          Socorro, NM 87801, USA \hfill}

\line{$^2$California Institute of Technology, Owens Valley Radio
          Observatory 105-24,\hfill}
\line{Pasadena, CA 91125, USA \hfill}

\bigskip
\hrule
\bigskip

\noindent{\bf Abstract.  Soft gamma-ray repeaters (SGR) are a class of
high energy transients whose brief emissions are thought to arise from
young\refto{Kulkarni1993} and highly magnetized neutron
stars\refto{Thompson1995, Thompson1996, Mazets1979a, Kouveliotou1998,
Hurley1998c}. The exact cause for these outbursts and the nature of
the energy loss remain unknown. Here we report the discovery of a
fading radio source within the localization of the relatively
under-studied SGR 1900+14. We argue that this radio source is a
short-lived nebula powered by the particles ejected during the intense
high energy activity in late August 1998, which included the
spectacular gamma-ray burst\refto{Cline1998} of August 27.  The radio
observations allow us to constrain the energy released in the form of
particles ejected during the burst, un-complicated by beaming effects.
Furthermore, thanks to the astrometric precision of radio
observations, we have finally localized this repeater to sub-arcsecond
accuracy.}

\bigskip
\bigskip
 \noindent{\it This manuscript has been submitted to Nature on 27
 October, 1998. For further enquiries please contact Dale Frail
(dfrail@nrao.edu) or Shri Kulkarni (srk@astro.caltech.edu).}

\vfill\eject

\refis{Thompson1995}
        Thompson, C., \& Duncan, R. C.
        The soft gamma repeaters as very strongly magnetized neutron
        stars - I. Radiative mechanism for outbursts.
        {\it M.N.R.A.S.} {\bf 275}, 255-300 (1995).

\refis{Thompson1996}
        Thompson, C., \& Duncan, R. C.
        The Soft Gamma Repeaters as Very Strongly Magnetized Neutron
        Stars. II. Quiescent Neutrino, X-Ray, and Alfven Wave Emission.
        {\it Astrophys. J.} {\bf 473}, 322-342 (1996).        

\refis{Kulkarni1993}
        Kulkarni, S. R. \&\ Frail, D. A.
        Identification of a supernova-remnant coincident with the soft
        gamma-ray repeater SGR 1806$-$20.
        {\it Nature} {\bf 365}, 33-35 (1993).

\refis{Mazets1979a}
        Mazets, E. P., Golenetskii, S. V., Il'inskii, V. N., Aptekar',
        R. L, \& Gur'yan, Yu. A. 
        Observations of a flaring X-ray pulsar in Dorado.
        {\it Nature} {\bf 282}, 587-589 (1979).

\refis{Kouveliotou1998} Kouveliotou, C. {\it et al.}  
        An X-ray pulsar with a super-strong magnetic field in the soft
        $\gamma$-ray repeater SGR 1806-20.
        {\it Nature} {\bf 393}, 235-237 (1998).

\refis{Hurley1998c}
        Hurley, K., Kouveliotou, C., Murakami, T., \&\ Strohmayer, T.
        {\it Intl. Astron. Union Circ. No.} {\bf 7001}, (1998).

\refis{Cline1998}
        Cline, T. L., Mazets, E. P., \& Golenetskii, S. V. 
        {\it Intl. Astron. Union Circ. No.} {\bf 7002}, (1998).


Until recently SGR 1900+14 was routinely labeled as the least prolific
of the SGRs with only three bursts\refto{Mazets1979b} in 1979 and
another three\refto{Kouveliotou1993} in 1992.  We investigated the
relatively crude localization obtained from the 1979 activity and
suggested\refto{Vasisht1994} that this SGR is associated with the
supernova remnant (SNR) \snr\ (see also ref. \Ref{Kouveliotou1994}).
Our Very Large Array (VLA) observations showed that \snr\ was a typical
shell-type object.  Furthermore, we suggested\refto{Vasisht1994} that
the X-ray source \xray\ located within the tighter 1994
localization\refto{Hurley1994} was the quiescent counterpart of the
SGR.  Subsequent ROSAT High Resolution Imager (HRI) observations
resulted in a tighter localization\refto{Hurley1996} of the source.

\refis{Mazets1979b}
        Mazets, E. P., Golenetskii, S. V., \& Gur'yan, Yu. A. 
        Soft gamma-ray bursts from the source B1900+14.
        {\it Sov. Asro. Lett.} {\bf 5}, 343-344 (1979).

\refis{Kouveliotou1993}
        Kouveliotou, C. {\it et al.}
        Recurrent burst activity from the soft gamma-ray repeater SGR
        1900+14.
        {\it Nature} {\bf 362}, 728-730 (1993).

\refis{Vasisht1994} 
        Vasisht, G., Kulkarni, S. R., Frail, D. A., \& Griener, G. 
        Supernova candidates for the soft $\gamma$-ray repeater 1900+14.
        {\it Astrophys. J.} {\bf 431}, L35-L38 (1994).      

\refis{Kouveliotou1994}
        Kouveliotou, C. et al. 
        The rarity of soft gamma-ray repeaters deduced from reactivation
        of SGR 1806$-$20.       
        {\it Nature} {\bf 368}, 125-127 (1994).   

\refis{Hurley1994}
        Hurley, K. {\it et al.}
        Network synthesis localization of two soft gamma repeaters.
        {\it Astrophys. J.} {\bf 431}, L31-L34 (1994).

\refis{Hurley1996}
        Hurley, K. {\it et al.}
        A possible X-ray counterpart to SGR 1900+14.
        {\it Astrophys. J.} {\bf 463}, L13-L16 (1996).


We began a monitoring program of \xray\ at the VLA motivated by the
notion that the high-energy bursts are most likely accompanied by
bursts of energetic particles and these particles would power a
synchrotron nebula, a mini-``plerion''.  Our initial observations of
the X-ray source showed no strong radio emission at or close to the
position of \xray.  In Table 1 we list the results from our VLA
monitoring study of \xray.

Our first VLA observation after the bright burst of August 27 was made
on September 3.  As can be seen from Figure 1, a new source is
apparent within the localization circle of \xray.  This radio source
was not seen in earlier VLA images obtained from data obtained at the
same frequency and similar resolution and sensitivity; see Figure 1.
The probability of finding a steady background radio source by
chance\refto{Windhorst1993} in this area is 10$^{-3}$. The unusual time
variability reduces the probability further still but by an unknown
factor. The radio source is unresolved and we place an upper limit to
the angular diameter of $0.8^{\prime\prime}$. The precise position of
the transient source (epoch J2000) is right ascension =
19h~07min~14.33s, declination =
+09\deg~19$^{\prime}$~20.1$^{\prime\prime}$, with an uncertainty of
0.15$^{\prime\prime}$ in each coordinate.

\refis{Windhorst1993}
        Windhorst, R. A., Fomalont, E. B., Partridge, R. B., \&
        Lowenthal, J. D.        
        {\it Astrophys. J.} {\bf 405}, 498-517 (1993).

%

Following the detection on September 3, the source underwent a rapid,
monotonic decay and is no longer detectable; see Table 1.  The
decaying portion of the curve can be fitted to a power law
(flux $\propto t^\delta$) with $\delta=-2.6\pm{1.5}$; here $t$ is 
time since the burst of August 27; see Figure 2.  On September 8
observations were also made at 1.43 GHz and 4.86 GHz, yielding a
spectral index $\alpha=-0.74\pm{0.15}$ (where the flux at frequency
$\nu$, S$_\nu\propto\nu^\alpha$); see Figure 2.  No linear or circular
polarization is detected but due to the faintness of the source, the
limits are not restrictive; the percentage polarization is $<30\%$.


Radio sources which show such transient behavior on this timescale are
quite rare in the sky. Furthermore, the source lies in the
Ulysses--CGRO triangulation\refto{Hurley1998c}, which has a total area
of 17 square arcminutes.  We suggest that the transient radio source
\vla\ is powered by the heightened burst activity from \sgr\ in late
August, which included the intense burst of August 27. If we are
correct, then we have localized \sgr\ to sub-arcsecond accuracy.
Hurley et al.\refto{Hurley1998c} report 5.17-s X-ray pulsations from
the vicinity of \xray. Such long period
pulsations\refto{Kouveliotou1998} are also seen from the
quiescent\refto{Murakami1994} counterpart of SGR 1806-20 and in the
bright burst\refto{Mazets1979a} from SGR 0526$-$66.  Thus \xray\ is
very likely the X-ray counterpart of SGR 1900+14.

\refis{Murakami1994}
        Murakami, T. {\it et al.} 
        X-ray identification of the soft $\gamma$-ray repeater
        1806$-$20.
        {\it Nature} {\bf 368}, 127-129 (1994).


SGR 1806$-$20 offers an excellent analog against which we now
interpret our radio observations of \sgr. To start with, both SGRs are
associated with SNRs: SGR 1806$-$20 is embedded in the plerionic SNR
G10.0$-$0.3 (refs. \Ref{Kulkarni1994, Vasisht1995, Frail1997}) whereas
SGR 1900+14 is found just outside \snr\ (ref. \Ref{Vasisht1994}).  The
intense bursts, the quiescent X-ray counterparts, the long-period
pulsations and the associated supernova remnants have been interpreted
in the framework of the magnetar
model\refto{Thompson1995,Thompson1996}.  Magnetars are highly
magnetized neutron stars with dipole field strengths of
$10^{14}-10^{15}$ G, considerably larger than those of radio pulsars.
The ultimate source of energy for the bursts and the quiescent
emission comes from the decay of the magnetic field. The non-thermal
quiescent X-ray emission and the highly suggestive radio
images\refto{Vasisht1995,Frail1997} of SGR 1806$-$20 provide
compelling evidence for a steady particle wind from the SGR.  The
``nested'' appearance\refto{Kulkarni1994} of G10.0$-$0.3 is best
explained\refto{Vasisht1995} as being powered by an episodic injection
of energy from the underlying SGR.  Thus G10.0$-$0.3 is powered both
by a steady and episodic power source.

\refis{Kulkarni1994} 
        Kulkarni, S. R., Frail, D. A., Kassim, N. E., Murakami, T. \&
        Vasisht, G. 
        The radio nebula of the soft $\gamma$-ray repeater 1806$-$20.
        {\it Nature} {\bf 368}, 129-131 (1994).

\refis{Vasisht1995}
        Vasisht, G., Frail, D. A., \& Kulkarni, S. R. 
        Radio monitoring and high-resolution imaging of the soft
        gamma-ray repeater 1806-20.
        {\it Astrophys. J.} {\bf 440}, L65-L68 (1995).        

\refis{Frail1997}
        Frail, D. A., Vasisht, G., \& Kulkarni, S. R. 
        The Changing Structure of the Radio Nebula around the Soft
        Gamma-Ray Repeater SGR 1806$-$20.
        {\it Astrophys. J.} {\bf 480}, L129-L132 (1997).



\vla\ shares characteristics similar to those of G10.0$-$0.3.  Both
have low brightness temperatures (a lower limit of 10 K is derived for
\sgr\ from the upper limit on the angular size of
0.8$^{\prime\prime}$) and unusual (for a plerion) non-thermal radio
spectra indices ($-0.6$ and $-0.74$).  The main distinction is that
G10.0$-$0.3 is bright and long-lived whereas \vla\ is short-lived and
fainter.  We attribute these differences to (1) the low-pressure
environment surrounding SGR 1900+14 and (2) the lower energy loss and
activity of SGR 1900+14 as compared to that of SGR 1806$-$20.  The
radio nebula around SGR 1806$-$20 is bright because this prolific SGR
(with implied large energy loss) is embedded in a high pressure
environment, the supernova remnant G10.0$-$0.3.  Thus the particle
wind emanating from SGR 1806$-$20 is contained, thereby accounting for
the high surface brightness. In contrast, SGR 1900+14 is not as
prolific and lies outside G\thinspace{42.8+0.6}. Thus the particle
wind from SGR 1900+14 is confined only by the ram pressure from the
motion of SGR 1900+14 through interstellar medium. This weaker
confinement does not allow for the buildup of a plerion, thereby
resulting in a weaker plerion.

Accepting our reasoning that \vla\ is a synchrotron emitting nebula we
apply the synchrotron model\refto{Pacholczyk1970} to derive physical
parameters. There are two major unknowns: the distance and angular
size of the source. On general grounds, the distance to \sgr\ is
roughly 10 kpc, given that it is located in the inner Galaxy.
However, if SGR 1900+14 is associated with \snr\ then a distance of 5
kpc is probably reasonable\refto{Vasisht1994}. The minimum energy of the
nebula (integrating the spectrum shown in Figure 2 from $10^7$ Hz to
$10^{11}$ Hz) and the equipartition magnetic field strength is then
$U_{\rm min} =3\times 10^{42} d_5^{17/7}\theta_4^{9/7}\, {\rm erg}$ and
$B_{\rm min} = 0.55 d_5^{-2/7} \theta_4^{-6/7}\, {\rm mG;}$ here the
distance is $5d_5$ kpc and the angular radius of the nebula is 
$\theta=0.4\theta_4$
arcseconds.

Using the formulation of Scott \&\ Readhead\refto{Scott77} we can
obtain a robust estimate of the angular size of the source, the
so-called ``equipartition radius'', $\theta_{\rm eq}= 165\,
d_5^{-1/17} S_p^{8/17}\nu_p^{-1.07}$ $\mu$arcsecond where $\nu_p$ is
the peak of the synchrotron spectrum (in GHz) and $S_p$ the
corresponding flux (in mJy).  The total energy of a source of size
$\theta$ is $U=1/2U_{\rm eq}\eta^{11}(1+\eta^{-17})$ where
$\eta=\theta/\theta_{\rm eq}$ and $U_{\rm eq}$ is the minimum energy
of a source with an angular radius of $\theta_{\rm eq}$. Small
deviations from $\theta_{\rm eq}$ result in steep energy demands and
that is why most sources have angular sizes quite close to
$\theta_{\rm eq}$.  Unfortunately, we do not have sufficient
wavelength coverage to see the true synchrotron peak.  Instead we use
the lowest frequency data (1.43 GHz) on September 08 to obtain a true
lower limit $\theta > \theta_a \sim 100\mu$arcsecond.  Using this
angular radius results in $U_{\rm a}\sim 7\times{10}^{37}$ erg and
$B_a\sim 0.7$ G. The radiative decay timescale for the electrons
responsible for the 8-GHz emission is then about 2 years.

\refis{Pacholczyk1970}
        Pacholczyk, A. G. 1970, Radio Astrophysics, Chap. 7 (San
        Francisco: Freeman)

\refis{Scott77}
        Scott, M. A. \& Readhead, A. C. S.
        The low frequency structure of powerful radio sources and
        limits to departures from equipartition.
        {\it M.N.R.A.S.} {\bf 180}, 539-550 (1977).

Our incomplete knowledge of the true angular size of \vla\ does not
allow us to pinpoint the region from which the radio electrons
originate and to infer the total particle energy and the magnetic
field.  We have two limits for $\theta$ and both limits are viable.
The strength of the magnetic field of a $6\times 10^{14}$ G magnetar
rotating at 5.16 s (ref. \Ref{Hurley1998c}) at the edge of the radio
nebula of angular radius $\theta$ is $B_{\rm M} \sim 0.13
(\theta/\theta_a)^{-1}d_5^{-1}$ G, comparable to $B_a$ if $\theta\sim
\theta_a$.  In the other extreme, when $\theta=0.4$ arcseconds, the
magnetic field from the magnetar would be negligible but the inferred
field strength can be easily explained as arising from compression of
the ambient field by the bow shock. The mean expansion speed of the
nebula is $3.5\times 10^{10} d_5\theta_4t_{10}^{-1}$ cm s$^{-1}$ where
$t_{10}$ is the age of the nebula in units of 10 days.  Thus in the
limit of $\theta=0.4$ arcseconds the nebula would have expanded
relativistically. This is possible if previous outbursts have swept up
the ambient gas.  Interestingly enough, in this limit, $U_{\rm min}$
is comparable with the isotropic burst energy of $3\times
10^{42}d_5^2$ erg, as estimated from the fluence of the August 27
burst ($\sim{10}^{-3}$ erg cm$^{-2}$, M.~Feroci, pers. comm.).  In
either case, it is clear that nebular expansion accounts for the
rapidly decaying radio emission. Thus the above energy estimates
derived from observations 7--10 days after the burst need to be
significantly revised upwards to obtain the initial release of energy.

In the future the availability of broad-band observations and/or
direct measurement of the size of the nebula will enable us to
directly estimate the energy of the burst in particles (without the
uncertainties of beaming that bedevil estimates from gamma-ray data).
Equally important are accurate measurements of the average particle
luminosity, since particle-aided spindown can substantially modify
estimates for the magnetic field and the characteristic age of a
neutron star\refto{Blaes1998}. These estimates, to our knowledge, are
unobtainable in any other fashion.  In those cases where the SGR is
immersed in a high pressure region (e.g. SGR 1806$-$20) we are able to
trace the entire history of energy loss from the magnetar. The
sub-arcsecond localization presented in this paper will greatly help
in identifying possible stellar counterparts of this SGR (as was done
for SGR 1806$-$20; refs.~\Ref{Kulkarni1995,vanKerkwijk1995}).


\refis{Blaes1998}
        Thompson, C. \& Blaes, O. 
        Magnetohydrodynamics in the extreme relativistic limit.
        {\it Phys. Rev. D} {57} 3219-3234, (1998).

\refis{Kulkarni1995}
        Kulkarni, S. R., Matthews, K., Neugebauer, G., Reid, I. N.,
        van Kerkwijk, M. H. \& Vasisht, G.
        Optical and infrared observations of SGR 1806$-$20.
        {\it Astrophys. J.} {\bf 440}, L61-L64 (1995).        
    
\refis{vanKerkwijk1995}
        van Kerkwijk, M. H., Kulkarni, S. R., Matthews, K., \&
        Neugebauer, G. 
        A luminous companion to SGR 1806$-$20.
        {\it Astrophys. J.} {\bf 444}, L33-L35 (1995).        





\bigskip
\centerline {\bf References}
\bigskip
\endreferences

\noindent{\bf Acknowledgments.}

The National Radio Astronomy Observatory is a facility of the National
Science Foundation operated under cooperative agreement by Associated
Universities, Inc.  DAF thanks C. Thompson and M. Goss for useful
discussions. We thank B. Clark and M. Goss for their long-term support
in the search for \sgr.  SRK's research is supported by the National
Science Foundation and the National Aeronautics \&\ Space
Administration.

\centerline{\hbox{\psfig{file=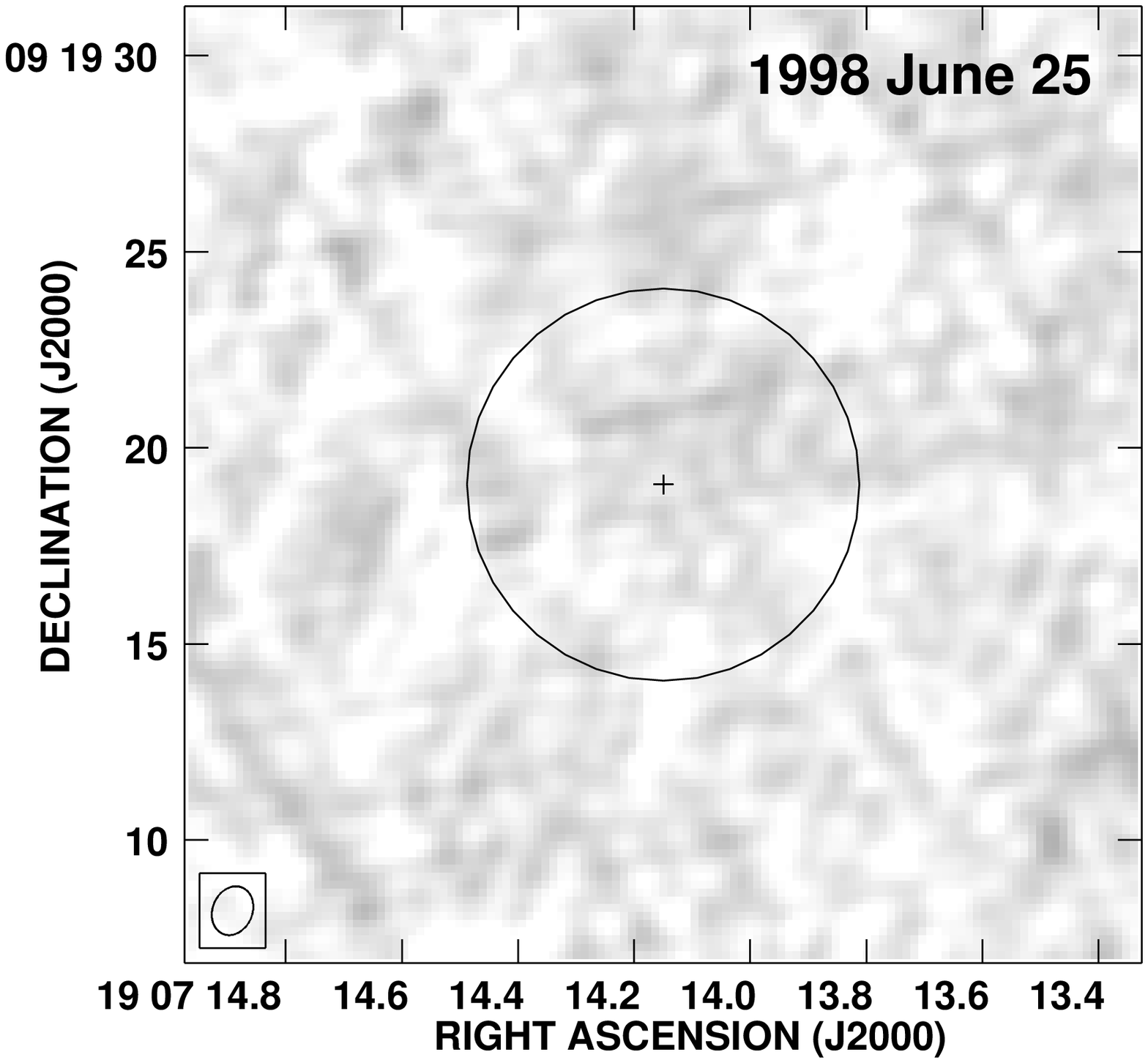,width=3.4in,clip=}} \hskip
-0.3in \hbox{\psfig{file=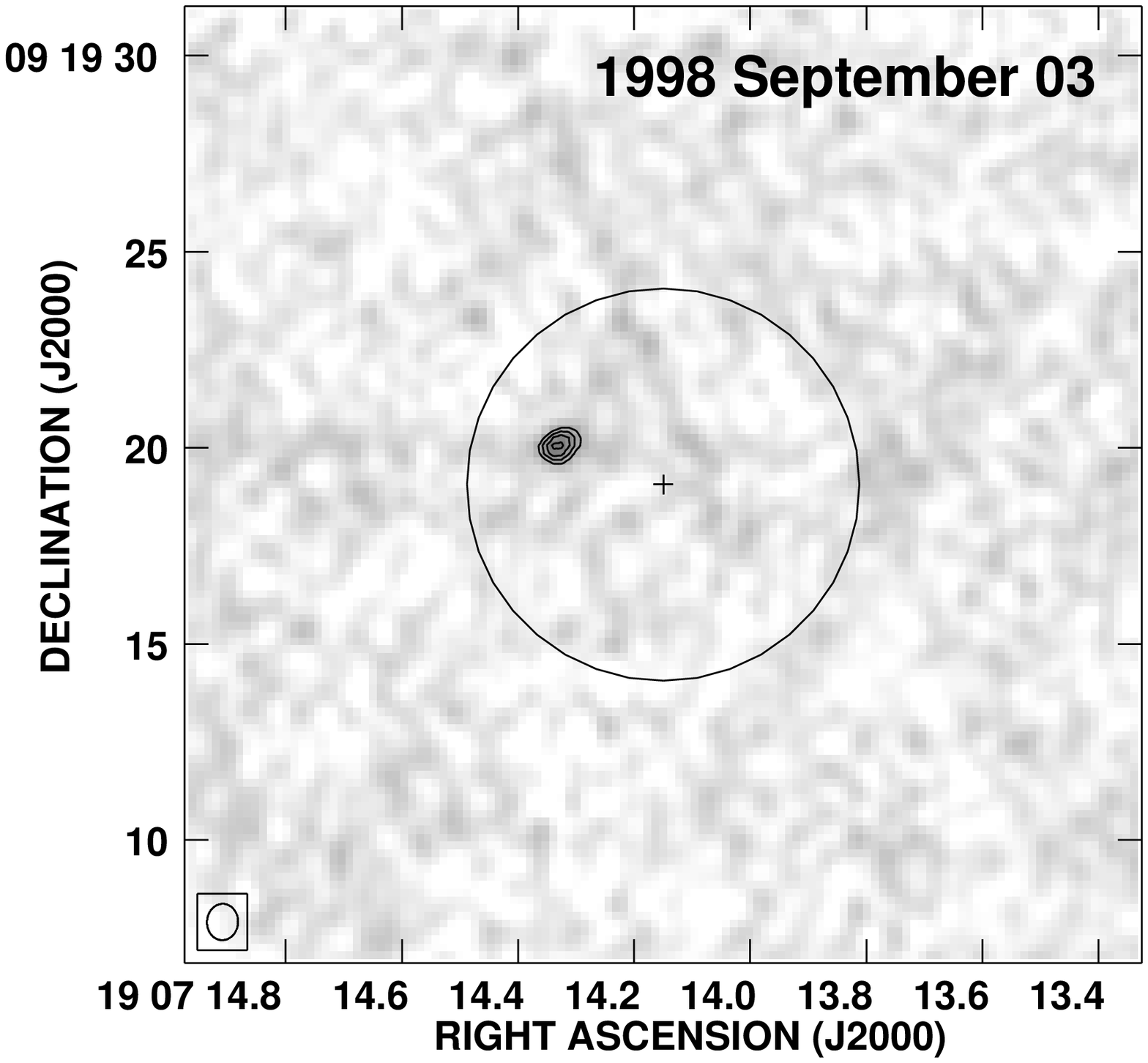,width=3.4in,clip=}}}


\noindent Figure \localization.  {\eightrm Images made at 8.46 GHz of
the field around \xray, a source proposed\refto{Vasisht1994} to be the
X-ray counterpart of \sgr.  Although ref. \Ref{Hurley1996} does not
discuss the size of the error circle, our assumed value of
10-arcsecond radius is an exceedingly conservative value. The ``+''
sign is the nominal location of \xray.  The June 25 image does not
show any evidence for a radio source to a 2-$\sigma$ level of 92
$\mu$Jy. On September 3 there is an unresolved 300 $\mu$Jy radio
source \vla\ within the X-ray localization. This source faded to an
undetectable level by October 1.}

\vfill\eject
\centerline{\hbox{\psfig{file=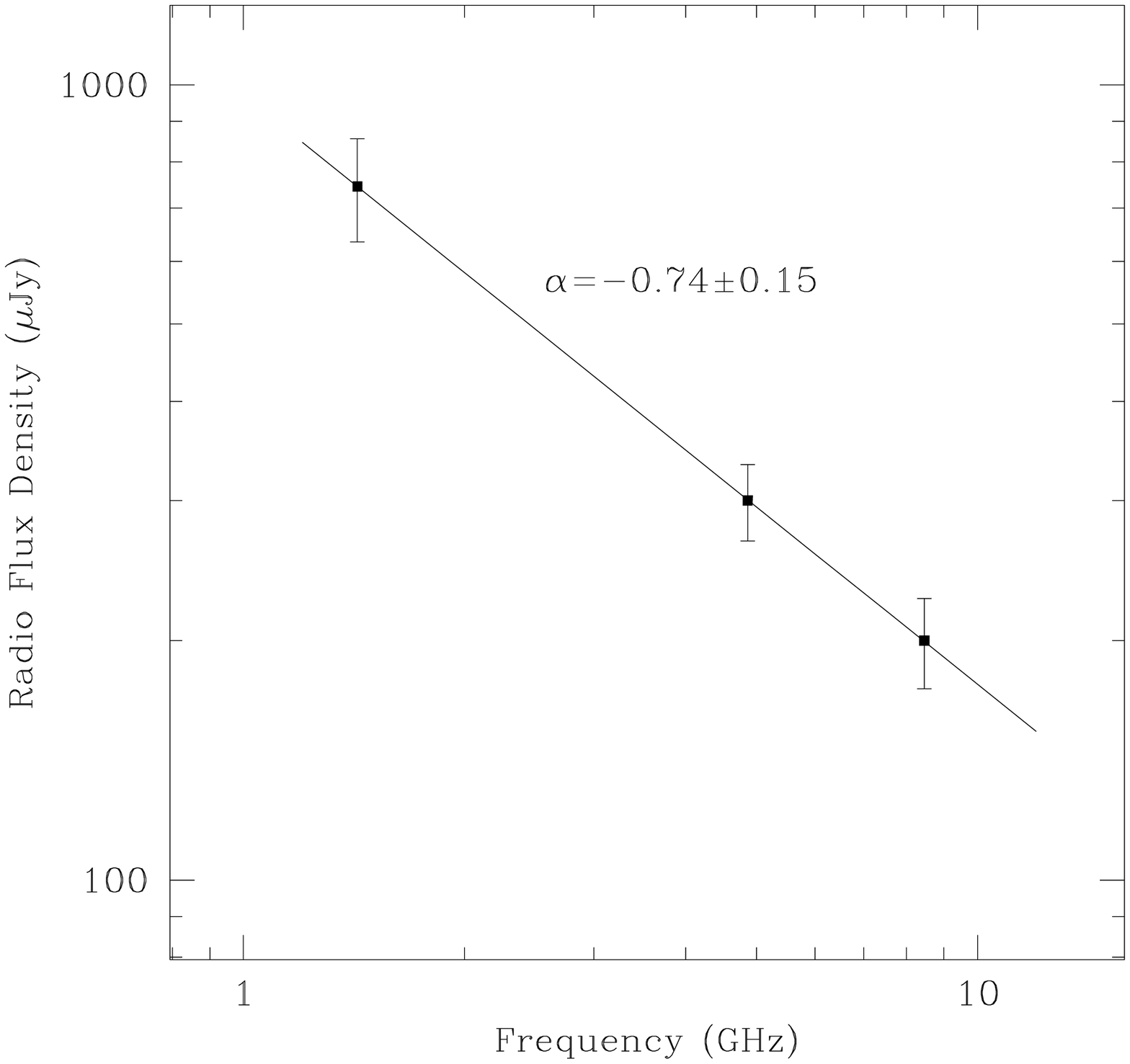,width=3.5in,clip=}} \hskip
-0.3in
\hbox{\psfig{file=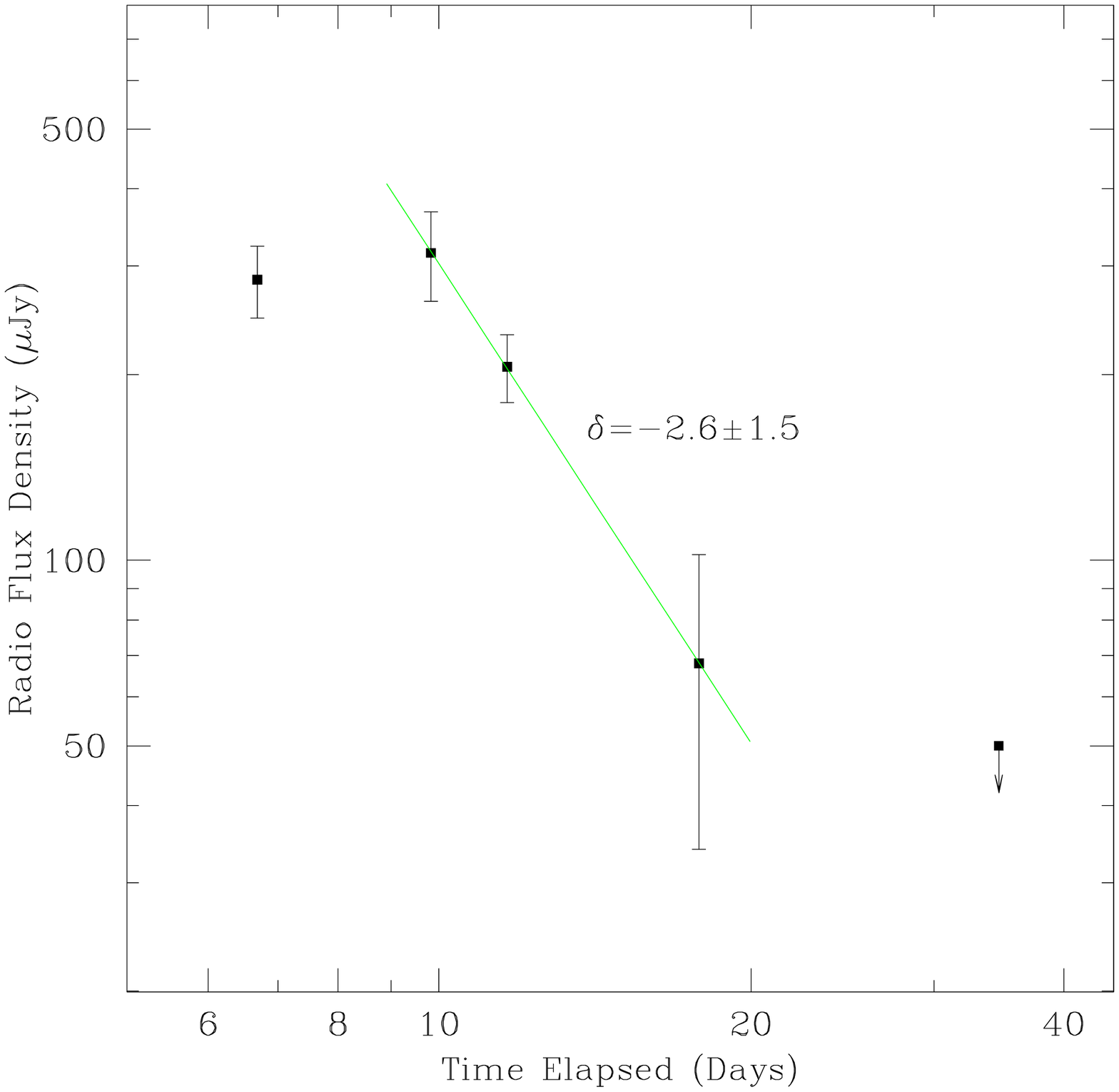,width=3.4in,clip=}}}

{\eightrm
\noindent Figure \radio. (left) The spectrum of the radio transient on
  1998 September 8 between 1 and 10 GHz. A weighted least-squares fit
  to these data gives a spectral slope $\alpha=-0.74\pm${0.15}.
  (right) The light curve at 8.46 GHz from 1998 September 3 to October
  10. Since the exact time for the onset of the radio emission is
  unknown, the horizontal axis is drawn assuming that it originated at
  the same time as the gamma-ray burst of August 27.432 UT. A weighted
  least-squares fit to the declining portion of the light curves gives
  a power-law slope of $\delta=-2.6\pm{1.5}$.
}

\vfill\eject
\topinsert{
$$
\table
\tablespec{\l\c\c\c\c}
\body{
\header{\bf Table \fluxtable. VLA Observations of
RX\thinspace{J190717+0919.3}}
\skip{15pt}
\hline
\skip{5pt}
Date & Freq. & Flux    & rms     \end
(UT) & (GHz) & ($\mu$Jy) & ($\mu$Jy) \end
\skip{5pt}
\hline
\skip{5pt}
1994 Oct. 05.00 & 1.43 & \omit & 110  \end
1994 Oct. 05.01 & 8.41 & \omit &  30  \end
1995 Dec. 14.80 & 8.41 & \omit &  33  \end
1995 Dec. 21.96 & 8.41 & \omit &  55  \end
1995 Dec. 26.81 & 1.43 & \omit & 125  \end
1995 Dec. 26.82 & 4.59 & \omit &  44  \end
1998 Jun. 25.25 & 8.46 & \omit &  46  \end
1998 Sep. 03.12 & 8.46 &   285 &  38  \end
1998 Sep. 06.26 & 8.46 &   315 &  52  \end
1998 Sep. 08.07 & 1.43 &   745 & 110  \end
1998 Sep. 08.08 & 4.86 &   300 &  33  \end
1998 Sep. 08.09 & 8.46 &   206 &  26  \end
1998 Sep. 14.24 & 8.46 &    70 &  43  \end
1998 Oct. 01.06 & 8.46 & \omit &  25  \end
1998 Oct. 11.12 & 1.43 & \omit &  70  \end
\skip{5pt}
\hline
}
\endtable
$$
}\endinsert

\par

{\eightrm
\noindent
(a) The entries (from left to right): the UT date of the observation,
the observing frequency (in GHz), the flux density of the source if it 
was detected, and the rms noise in the image. 

\noindent (b) All observations were made with a bandwidth of 100 MHz
at each frequency. Antenna phase calibration was accomplished
using extragalactic radio sources with well-known positions near \sgr.
The absolute flux scale (accurate to better than $\pm$2\%) at each
epoch was fixed by short observations of the radio sources
3C\thinspace{48}, 3C\thinspace{147}, or 3C\thinspace{286}. 

\noindent (c) The angular resolution at 8.46 GHz was approximately
0.8--1.2 arcsecond, increasing linearly with decreasing frequency.
}

\bye